# Unified Entropy-Ruled Einstein's Relation for Bulk and Low-Dimensional Systems: A Hopping to Band Shift Analysis

K. Navamani

*Department of Physics*
*Centre for Research and Development (CFRD)*
*KPR Institute of Engineering and Technology*
*Coimbatore-641407, India*
*E-mail: pranavam5q@gmail.com*

In this letter, we present the unified paradigm on entropy-ruled Einstein's diffusion-mobility relation (µ/D ratio) for all dimensional systems (1D, 2D and 3D) of molecules and materials. The different dimension-associated fractional value of the variation in differential entropy with respect to the chemical potential ($\Delta h_S/\Delta \eta$) gives the quantum-classical transition version of $\mu/D$ relation. This is a new alternative version for quantum devices, instead of Einstein's original relation of $\mu/D = q/k_BT$; where $q$, $k_B$ and $T$ are the electric charge, Boltzmann constant and temperature, respectively. It is found that the fractional value of $\Delta h_S/\Delta \eta$ for µ/D ratio for different dimensional systems or devices is a direct consequences with the average energy ($\bar{E}$)-Fermi energy ($E_F$) relation, which can varies with the typical dimensions, whether the system belongs to 1D or 2D or 3D. This unified entropy-ruled transport formalism works well for both the quantum and classical systems with equilibrium as well as non-equilibrium conditions. Based on the dimensional dependent entropy-ruled µ/D factor, the Navamani-Shockley diode equation is transformed.

The diffusion-mobility relation is the fundamental transport relation to study any semiconducting materials [1,2]. Over a century, the Einstein relation ($D/\mu = k_BT/q$) has been used to estimate the basic electronic property for disordered systems (e.g., molecules). This classical relation is more suitable at high temperature regime (T≥150K) for both the equilibrium and quasi-equilibrium conditions [1,3]. Based on the numerous studies, it has been observed that for degenerate devices, the Einstein relation does not quantify the exact electronic property via the diffusion-based mobility calculations [3-7]. In real life, the high performance electronic devices are principally designed by degenerate materials [4]. The electronic characteristic behaviour of these functional devices can be tuned by the appropriate doping or substitutions. Hence, the quantum level of understanding in molecular/material systems for both the cases of equilibrium and nonequilibrium is crucial to design the advanced functional devices. In this regard, the exact diffusion-mobility relation is highly solicited, instead of classical Einstein relation, to explore the electronic property of wide ranges of systems, from molecular to material systems with different dimensions (quantum wires, layered thin films and degenerate bulk system). Moreover, most of the recent studies and research reports urge the necessity of revisiting or new development in the diffusion-mobility relation for next generation semiconductor technology [2,3,6-8]. For degenerate (electron/hole rich) semiconducting devices, the delocalization of charge density is a key parameter to functionalize the electronic behaviour. The charge distribution in the electronic states is a direct controlling factor for any semiclassical and quantum devices. Thus, here the understanding of charge density ($n$) and density of states (DOS) in the materials are the crucial parameters for quantum transport, which is directly related with $\mu/D$ relation [3,4,7].

To get the complete picture of charge transport mechanisms in different dimensional electronic systems, here we present the unified formalism of entropy-dependent carrier density

Entropy-Ruled Method

for 1D, 2D and 3D as,

$$n_{h_S,S}\big|_{1D} = n_0\big|_{1D} \exp\left(\frac{1}{3}\left(h_S - \frac{S}{k_B}\right)\right) \quad (1)$$

$$n_{h_S,S}\big|_{2D} = n_0\big|_{2D} \exp\left(\frac{1}{2}\left(h_S - \frac{S}{k_B}\right)\right) \quad (2)$$

$$n_{h_S,S}\big|_{3D} = n_0\big|_{3D} \exp\left(\frac{3}{5}\left(h_S - \frac{S}{k_B}\right)\right) \quad (3)$$

Here, $n_0$, $h_S$ and $S$ are the initial charge carrier density, differential entropy and thermal entropy, respectively. In principle, the differential entropy for whole range (i.e., limits start from $-\infty$ to $+\infty$) is obtained by the relation, $h_S(x) = -\int_{-\infty}^{+\infty}\Phi(x).\ln\Phi(x)dx = \ln(\sigma\sqrt{2\pi e})$, where $\Phi(x)$, $\sigma$ are the function and Gaussian disorder width (or Gaussian variance), respectively [3,8]. The Gaussian function is defined by $\frac{1}{\sigma\sqrt{2\pi}}\exp\left(-\frac{x^2}{2\sigma^2}\right)$. In this letter, the effective entropy $h_{eff} = h_S - S/k_B$. For degenerate systems/devices, it is expected that $h_{eff} \rightarrow h_S$, since $h_S \gg S$. As described in earlier studies, the generalized Einstein's $\mu/D$ relation can be governed by the ratio between the DOS ($=dn/d\eta$) and charge density, where $\eta$ is the chemical potential [3,4,7]. The generalized Einstein relation does work up to small perturbation conditions only; but not suitable for highly perturbed systems of molecules and materials. Through the entropy-dependent carrier density expressions for different dimensional systems, one can resolve the above issue in $\mu/D$ relation, which is the unified entropy-ruled Einstein relation [3,9]. This gives more accuracy in results for diffusion-based mobility transport calculation for any dimensional devices (e.g., quantum wires, layered devices) at wide thermodynamic situations. In this context, we derive the entropy-ruled Einstein's relation using Eqns. 1-3, and are expressed as,

$$\frac{\mu}{D}\bigg|_{1D} = \frac{q}{3}\frac{\Delta h_{eff}}{\Delta\eta} \quad (4)$$

$$\frac{\mu}{D}\bigg|_{2D} = \frac{q}{2}\frac{\Delta h_{eff}}{\Delta\eta} \quad (5)$$

$$\frac{\mu}{D}\bigg|_{3D} = \frac{3q}{5}\frac{\Delta h_{eff}}{\Delta\eta} \quad (6)$$

Based on the above relations (Eqns. 4-6), the unified entropy-ruled $\mu/D$ relation is defined as,

$$\left(\frac{\mu}{D}\right)_d = \left(\frac{d}{d+2}\right)q\frac{\Delta h_{eff}}{\Delta\eta} \quad (7)$$

Here, $d$ represents the dimension (1D, 2D, and 3D) of the systems. In this unified relation, the variation in chemical potential with respect to the effective entropy ($\Delta\eta/\Delta h_{eff}$) is referred to as the quantum-classical transition analogy [3]. It is observed that based on the dimensions, the fractional values ($d/(d+2)$) of $\Delta h_S/\Delta\eta$ are different, i.e., 1/3, 1/2 and 3/5 for 1D, 2D and 3D, respectively. These fractional values are principally associated with the relations $\frac{E}{N} = \frac{1}{3}E_F$ for 1D, $\frac{E}{N} = \frac{1}{2}E_F$ for 2D, and $\frac{E}{N} = \frac{3}{5}E_F$ for 3D systems, respectively [10,11]. That is, the dimension dependent fractional values directly involve the carrier density-entropy relations (see Eqns. 1-3). The degeneracy strength of the materials can be quantified by $h_S$ via the Gaussian width ($\sigma$), since $h_S = \ln(\sigma\sqrt{2\pi e})$. The formation of degeneracy levels can be tuned by the applied electric or magnetic field.

In principle, the charge transport in molecular solids characterized by the Marcus theory of charge transfer rate [8,12],

$$k = \frac{J_{eff}^2}{\hbar}\sqrt{\frac{\pi}{\lambda k_B T}}\exp\left[-\frac{(\Delta E_{ij} - \lambda)^2}{4\lambda k_B T}\right] \quad (8)$$

where, $J_{eff}$, $\hbar$, $\lambda$ and $\Delta E_{ij}$ are effective charge transfer integral, reduced Planck constant, reorganization energy, and site energy difference between adjacent electronic sites of neighboring molecular units, respectively. In this study, $\Delta E_{ij} = \varepsilon_j - \varepsilon_i + q\vec{E}.\vec{R}_{ij}$, where $\vec{E}$ and $R_{ij}$ are the applied electric field and stacking distance. The site energy difference can be tuned by the applied electric field or appropriate doping, which improves the charge transfer rate; accordingly other charge transport quantities are increased.

The Marcus theory of rate Eqn. 8 is the Gaussian type function, which gives normal diffusion transport [8]. In this work, the vibronic contribution (i.e., $\lambda$) to the Gaussian width ($\sigma_v$) is described by $\sigma_v = \sqrt{2\lambda/(k_B T)}$, and hence the expected vibronic entropy is explicitly expressed as $h_v(\lambda) = \ln(\sigma_v\sqrt{2\pi e}) = \frac{1}{2}\ln(4\lambda\pi e/(k_B T))$ [13,14]. On the other hand of electric field-assisted carrier's energy flux contribution to the differential entropy, $h_S$, is quantified by the relation [8,9],

$$h_S(\vec{E}) = \ln\left(\frac{dE(\vec{E})/dt}{dE(\vec{E}=0)/dt}\right) \quad (9)$$

Here, $dE(\vec{E})/dt$ and $dE(\vec{E}=0)/dt$ are the rate of traversing energy (during the charge transport) along the hopping sites with and without applied electric field. Now the actual or effective differential entropy is $h_A = h_{eff} \equiv h_S(\vec{E}) - h_v(\lambda)$. The vibronic contribution (reorganization energy) in entropy term resists the electron/hole transfer kinetics, but the carrier's energy flux (rate of traversing energy) due to electric field favours the transport. At $h_S(\vec{E}) \gg h_v(\lambda)$ limits, the large degeneracy is expected even at small applied electric or magnetic fields, which improves the mobility. On the other hand $(h_S(\vec{E}) < h_v(\lambda))$, the localized transport is anticipated in the studied molecular solids and hence the required bias (in the forms of gate-voltage, electric field, doping, etc.) will be more to activate the electron/hole dynamics. Here, the parameter $h_S$ increases with the electronic coupling, $J_{eff}$. According to the unified entropy-ruled μ/D relation (Eqn. 7), the main observation is that the differential entropy directly influences the mobility, and the chemical potential is responsible for the enhancement in diffusion coefficient. Here, the term $\Delta\eta/\Delta h_S$ in the D/μ relation is also called the quantum-classical transition analogy, which describes the degeneracy stabilization potential [3]. The degeneracy stabilization potential is defined as the required minimum energy to maintain the existing degeneracy states by applied electric or magnetic field at finite temperature. For large relaxation or scattering potential (i.e., $\lambda > J_{eff}$), it is expected that the required degeneracy stabilization energy (or relative chemical potential over the entropy changes) will be large enough [3]. Through this analogy, one can explore the transformation from hopping to band, and vice-versa at appropriate thermodynamic conditions. Hence, the methodology for measuring the chemical potential is important for different physical systems such as molecules and materials.

For molecular species, the required chemical potential for the charge transfer kinetics is defined as the barrier height of the potential landscape of an adjacent sites, which is naturally associated with the Marcus theory [3,15]

$$\eta = k_B T \left(1 + \ln\left(\frac{n}{n_o}\zeta\right)\right) \quad (10)$$

Here, $\zeta$ is the activity coefficient, which is represented by

$\zeta = \exp\left[\frac{(\lambda + \Delta E(\vec{E}))^2}{4\lambda k_B T}\right]$, and hence the generalized chemical potential (compare Eqns. 1-3 and 10) for different dimensional electronic systems

$$\eta_d = k_B T \left(1 + \left(\frac{d}{d+2}\right) h_{eff} + \frac{(\lambda + \Delta E(\vec{E}))^2}{4\lambda k_B T}\right) \quad (11)$$

By taking the differentiation with respect to the effective differential entropy, we get the chemical potential flux along the consequential hopping sites are expressed as,

$$\left.\frac{d\eta}{dh_{eff}}\right|_d = k_B T \left(\left(\frac{d}{d+2}\right) + \frac{(\lambda + \Delta E(\vec{E}))}{2\lambda k_B T}\frac{dE(\vec{E})}{dh_{eff}}\right) \quad (12)$$

Inserting Eqn. (12) in to Eqn. (7), we attain the unified entropy-ruled Einstein relation for 1D, 2D and 3D systems

$$\left.\frac{\mu}{D}\right|_d = \left(\frac{d}{d+2}\right)\frac{q}{k_B T}\frac{1}{\left[\left(\frac{d}{d+2}\right) + \frac{(\lambda + \Delta E(\vec{E}))}{2\lambda k_B T}\frac{dE(\vec{E})}{dh_{eff}}\right]}$$
(13)

This unified entropy-ruled μ/D relation is the new alternative version for molecular quantum devices, instead of Einstein's classical relation. This entropy-ruled relation works well for both the equilibrium and nonequilibrium conditions. Here, it is important to note that we retain the original Einstein relation again, while the energy flux along the electronic sites is zero, i.e., $\frac{dE(\vec{E})}{dh_{eff}} \to 0$ (see Eqn. 13). On the other hand, with respect to the magnitude of $\frac{dE(\vec{E})}{dh_{eff}}$, one can categorize the typical transport; whether it will be dominated by the diffusion coefficient or dominated by mobility. In this study, the energy flux variation due to the differential entropy will be $0 \leq \frac{dE(\vec{E})}{dh_{eff}} \leq 1$. In general, the reorganization energy (due to the presence of excess charge) has a direct relationship with the geometric structure, either linear or planar or 3D structure [12,16]. For molecular transport, the general formula of diffusion coefficient for organic semiconductors is

$$D = \frac{1}{2d}\frac{\Delta\langle R^2(t)\rangle}{\Delta t}, \quad (14)$$

where, $\langle R^2(t)\rangle$ is the mean squared displacement, $t$ is the simulation time, and $d$ represents the dimension (i.e., 1D, 2D and 3D) of the molecular crystals.

In this report, our unified entropy-ruled Einstein relation is verified for thiazolothiazole molecular derivatives **1** and **2** at different electric field-assisted site-energy disorder values. The chemical structure-based property relation of these derivatives was given in the earlier study [17]. It was observed that both thiazolothiazole derivatives **1** and **2** have good n-type performance (i.e., good electron mobility), and one can easily prepare in 54 and 18% yields, respectively [18]. Through end absorptions study, the obtained HOMO-LUMO gaps of these molecules were 3.06 and 2.48 eV, which showed the semiconducting properties of these molecules [17,18]. Moreover, the X-ray structure analyses revealed the stacking nature of these thiazolothiazole derivatives **1** and **2**. With these motivations, we are interested in doing the mobility calculation extensively. In this regard, the computed charge transport basic parameters (charge transfer integral, site energy and reorganization energy), and information about the molecular conformation of these derivatives are chosen from earlier reports, and we performed the numerical calculations to validate our entropy-ruled Einstein relation [9,17]. In this letter, the electric field-assisted site-energy differences $(\Delta E(\vec{E}))$ along with the vibronic coupled molecular relaxation (reorganization energy, $\lambda$) are included in our entropy-ruled charge transport calculation (see Eqn. 13) to quantify the exact diffusion-mobility for these thiazolothiazole derivatives **1** and **2**, which are shown in Figure 1. In the present study, the electron reorganization energy of these molecular derivatives **1** and **2** were 0.36 and 0.32 eV, respectively [17]. As discussed earlier, the effective differential entropy is defined as

$$h_{eff} = h_S(\vec{E}) - h_v(\lambda) \equiv \ln\left(\frac{dE(\vec{E})/dt}{dE(\vec{E}=0)/dt}\right) - \ln\sqrt{(4\lambda\pi e/(k_BT))}$$

Or,

$$h_{eff} = h_S(\vec{E}) - h_v(\lambda) \equiv \ln\left(\frac{dE(\vec{E})/dt}{dE(\vec{E}=0)/dt}\cdot\frac{1}{\sqrt{(4\lambda\pi e/(k_BT))}}\right) \quad (15)$$

Here, the first term describes the electric field effect on charge transport (via site-energy difference), which normally gives the drift effect on carrier dynamics. On the other hand, the normal mode of vibrations (in the form of $\lambda$) of the molecule resists the charge transfer process, which is represented by the second term in the above equation.

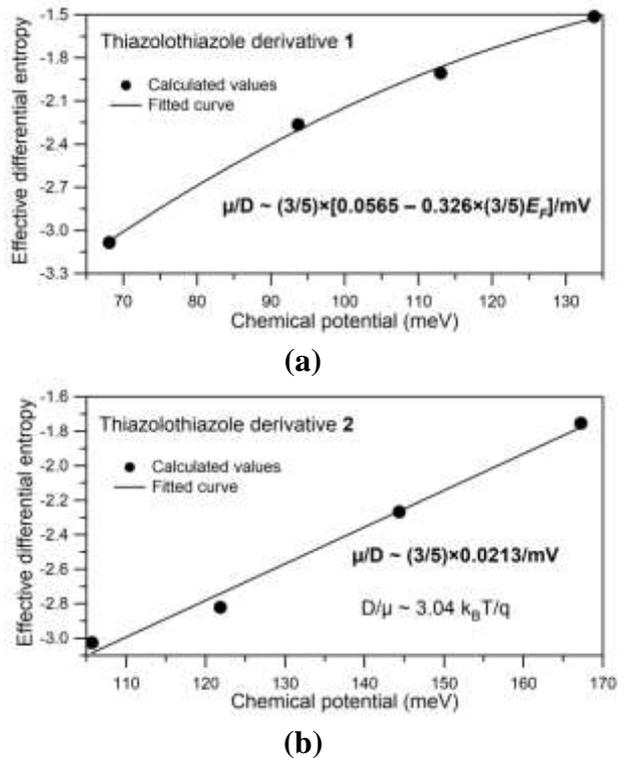

**(a)**

**(b)**

**Figure 1:** The effective differential entropy ($h_{eff}$) with respect to the chemical potential ($\eta$) for electron transport in thiazolothiazole derivatives (a) molecule **1** and (b) molecule **2** at different $\Delta E(\vec{E})$ values of 0, 25, 50 and 75 meV. For molecule **1**, the $h_{eff}$ is nonlinearly increasing with $\eta$; but for molecule **2**, the $h_{eff}$ increases with $\eta$ in linear manner. According to the shape of the $h_{eff}$-$\eta$ curve, the mobility-diffusion ($\mu/D$) relation is categorized; whether it follows Einstein's classical relation or it takes the deviation from Einstein's original value of $q/k_BT$. The electric field-assisted site-energy flux along the hopping sites is a deciding factor of transition between equilibrium to nonequilibrium transport or vice versa.

The rate of energy flux (i.e., rate of traversing carrier energy) with respect to the applied electric field in the molecular system can be characterized by energy redistribution relation. In principle, the carrier's energy flux has a direct relationship with the energy distribution shape in terms of width (here, Gaussian disorder width ($\sigma$)) in the concerned molecular solids. The logarithmic function of the Gaussian width is the differential entropy, $h_S = \ln(\sigma\sqrt{2\pi e})$. For degenerate systems, the expected disorder width will be more, and hence the magnitude of $h_S$ will be also large enough [3,9]. Noteworthy, the vibraonic contribution in the Gaussian width ($\sigma_v = \sqrt{2\lambda/(k_B T)}$) suppresses the weightage of degeneracy on electronic transport by the external interactions (e.g., Stark effect by electric field or Zeeman effect by magnetic field). At the same time, the electronic contributions in the Gaussian width ($\sigma$) improve the mobility via the degeneracy. Based on these consequences, we extend the numerical calculations for thiazolothiazole derivatives **1** and **2** at different set $\Delta E(\vec{E})$ values of 0, 25, 50 and 75 meV with fixed $\lambda$ and $T$ values. Using Eqns. 9, 11 and 15, the chemical potential ($\eta$) and effective differential entropy ($h_{eff}$) are calculated for these derivatives at different $\Delta E(\vec{E})$ values, and are plotted in Figure 1.

It is observed from Figure 1, the effective differential entropy increases in a nonlinear manner with respect to the chemical potential for the thiazolothiazole molecule **1**. On the other hand, for molecule **2**, the $h_{eff}$ linearly increase with $\eta$. The slope from $h_{eff}$-$\eta$ characteristic plot provides the exact $\mu/D$ factor. The large magnitude of $h_{eff}$ drives the mobility enhancement and the $\eta$ is responsible for enhancement in diffusion coefficient. The synergy between $h_{eff}$ and $\eta$ is the deterministic factor for $\mu/D$; accordingly, one can reveal the validity and limitations of the original Einstein relation. In this study, the changes in the energy flux by the differential entropy $\left(\dfrac{dE(\vec{E})}{dh_{eff}}\right)$ along the hopping sites decide the magnitude of $\mu/D$ values (Eqns. 12 and 13), which explicitly explains that whether the calculated exact diffusion-mobility value falls in Einstein region ($\mu/D = q/k_B T$) or not. The nature of enhancement in $h_{eff}$-$\eta$ plot typically categorizes the linear and nonlinear transport through the D/$\mu$ relation. Importantly, we retain the Einstein relation from our entropy-ruled D/$\mu$ relation under negligible or zero energy flux, $\dfrac{dE(\vec{E})}{dh_{eff}} \to 0$. Based on the amplitude of energy flux over the differential entropy tells how much far from equilibrium transport; that is, whether the D/$\mu$ follows equilibrium or quasi-equilibrium or nonequilibrium regime. For instance, both the molecules (thiazolothiazole derivatives **1** and **2**) have significant magnitudes of $\dfrac{dE(\vec{E})}{dh_{eff}}$, which confirms the nonequilibrium diffusion-mobility transport. Notably, for derivative **1**, the variation in $h_{eff}$ with respect to $\eta$, or vice-versa follows nonlinearly due to the asymmetric variation at different $\Delta E(\vec{E})$ values (i.e., non-uniform cooperative behaviour between $h_{eff}$ and $\eta$). On the other side, it is observed that for derivative **2** the change of $h_{eff}$ with respect to $\eta$ is linear, and hence the expected D/$\mu$ will be a constant value, which is shown in Figure 1 (b). In this report, the calculated D/$\mu$ value is around 78.25 mV, which is the equivalent to that of 3 times Einstein's original value of $k_B T/q$ (see Figure 1 (b)). Thus, the enhancement factor for this molecule **2** is around 3, which shows the deviation in Einstein relation. In other words, at each uniform interval of $\Delta E(\vec{E})$, the variation in diffusion coefficient (D) for derivative **2** is nearly 3 times larger than mobility ($\mu$). Here, the enhancement of D is larger than $\mu$ by applied electric field via $\Delta E(\vec{E})$. In this case, the required chemical potential is large for electron hopping over the barrier height. For thiazolothiazole derivative **1**, the enhancement in D is a non-uniform as well as in $\mu$, which leads to non-uniform variation in D/$\mu$ factor (see Figure 1 (a)). Based on the entropy-ruled $\mu/D$ relation, one can explicitly derive the other important charge transport quantities such as conductivity, current density, etc., for both the steady and nonsteady state conditions.

For different dimensional (1D, 2D and 3D) periodic materials, the average energy-Fermi energy relations follows, $\left.\frac{E}{N}\right|_{1D} = \frac{1}{3}E_F$, $\left.\frac{E}{N}\right|_{2D} = \frac{1}{2}E_F$, and $\left.\frac{E}{N}\right|_{3D} = \frac{3}{5}E_F$. In principle, the total energy per particle is the chemical potential ($\eta$) [19]. In this case, the total energy ($E$) comprises the kinetic, correlation and exchange terms, etc. Hence, the unified entropy-ruled diffusion-relation (see Eqns. 4-6 and 7) can be expressed as,

$$\frac{\mu}{D} = q\frac{\Delta h_{eff}}{\Delta E_F} \equiv q\frac{\Delta h_{eff}}{k_B \Delta T_F} \approx q\frac{1}{\sigma}\frac{d\sigma}{dE_F} \quad (16)$$

Since, $h_S = \ln(\sigma\sqrt{2\pi e})$. Here, $T_F$ is the Fermi temperature.
Or,

$$\frac{\mu}{D} \cong q\frac{(\sigma_f - \sigma_i)}{\sigma}\frac{1}{\Delta E_F} = q\frac{\Delta P}{\Delta E_F} \quad (17)$$

Here, $P$ is the probability of Gaussian disorder broadening (or width) due to Fermi energy (chemical potential) level shifting by external bias. For delocalized band transport, the electronic levels are very closely packed (i.e., continuum band) and hence the anticipated energy distribution is uniform [3]. This kind of intrinsic electronic structure typically leads to the inelastic scattering carrier dynamics. In general, for periodic materials, the electronic levels are aligned in the order of $\hbar^2 k^2 / 2m$; but for disordered molecular solids, the density of states are thermally ($k_B T$) populated, which is normally quantified by the Franck-Condon factor. In this regard, the above relation (see Eqns. 16 and 17, also see ref. 3) is simplified as,
(i) For periodic systems:

$$\frac{\mu}{D} = \frac{q}{E_F} = \frac{2q}{m^* v_F^2} \equiv \frac{2q\tau^2}{m^* l^2} \quad (18)$$

Here, $m^*$, $v_F$, $l$ and $\tau$ are the effective mass of an electron, Fermi velocity, mean free path and the relaxation time (or scattering time), respectively.
(ii) For disordered systems:

$$\frac{\mu}{D} = \frac{q}{k_B T} \quad (19)$$

To prove the validity of our entropy-ruled transport relation, here we have expressed the chemical potential for the 2D-Fermionic systems as,

$$\eta = k_B T \left\{ \ln\left[ \exp\left(\frac{\pi\hbar^2}{m^* k_B T} n_{2D}\right) - 1 \right] \right\} \quad (20)$$

Using Eqn. 20, the entropy-ruled D/μ relation for 2D systems (see Eqns. 5, 7) is given by

$$\left(\frac{D}{\mu}\right)_{2D} = \frac{2}{q}\frac{d\eta}{dh_{eff}} = \frac{E_F}{q}\left[\frac{\exp\left(\frac{\pi\hbar^2}{m^* k_B T} n_{2D}\right)}{\exp\left(\frac{\pi\hbar^2}{m^* k_B T} n_{2D}\right) - 1}\right] \quad (21)$$

For degenerate systems of molecules and materials, the above relation becomes,

$$\left(\frac{D}{\mu}\right)_{2D} = \frac{2}{q}\frac{d\eta}{dh_{eff}} = \frac{E_F}{q} \quad (22)$$

Or

$$\left(\frac{\mu}{D}\right)_{2D} = \frac{q}{2}\frac{dh_{eff}}{d\eta} = \frac{q}{E_F} = \frac{2q\tau^2}{m^* l^2} \quad (23)$$

This is the same relation as shown in Eqn. 18. In such a way, we can also attain the similar D/μ relation for degenerate quantum 1D and 3D systems. On the other hand, for disordered molecular solids, the D/μ relation at thermal equilibrium cases is $q/k_B T$. In the present study, we have revisited the Boltzmann approach using entropy-ruled method as [3,7],

$$\mu = \frac{q(dn/d\eta)v_F^2 \tau}{2n} \quad (24)$$

In this context, the entropy-ruled Boltzmann treatment of mobility for 1D, 2D and 3D materials are formulated as,

$$\mu_d = \left[\left(\frac{d}{d+2}\right)q\frac{dh_{eff}}{d\eta}\right]\frac{v_F^2 \tau}{2} \quad (25)$$

Comparing Eqn. 25 with Eqn. 7, the diffusion coefficient can be explicitly defined as,

$$D = \frac{v_F^2 \tau}{2} \equiv \frac{l^2}{2\tau} \quad (26)$$

Substituting Eqn. 26 into Eqn. 23, we get the mobility

$$\mu = \frac{q\tau}{m^*} \quad (27)$$

This is the generalized mobility for Schrödinger-type degenerate quantum systems. For Dirac materials, the mobility is transformed as,

$$\mu = \frac{q}{\hbar k}\frac{v_F \tau}{2} = \frac{q}{\hbar k}\frac{l}{2} \quad (28)$$

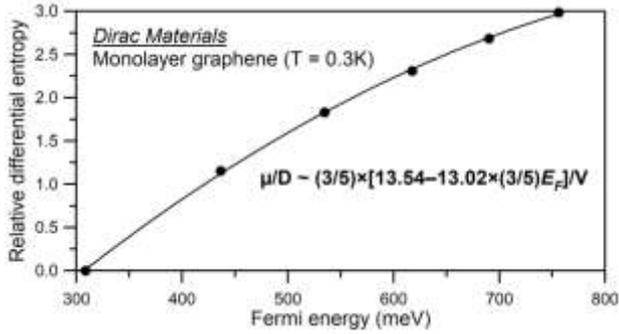

**Figure 2:** The calculated differential entropy with respect to the gate-voltage-dependent Fermi energy (or chemical potential) in the monolayer graphene. The electron density flux is here characterized by diffusion-based mobility relation via the entropy-ruled method. The ratio between the relative differential entropy and Fermi energy is a direct consequence of the µ/D ratio.

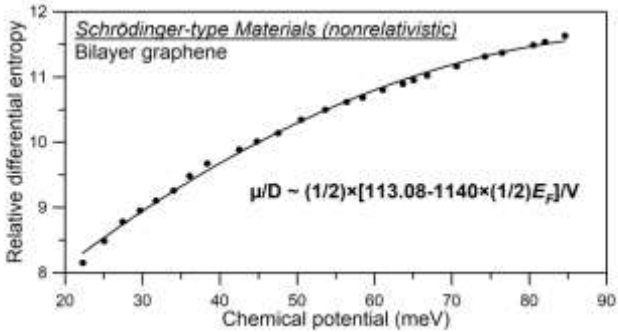

**Figure 3:** The calculated relative differential entropy at different chemical potential with T=1.4K for the sample 2 bilayer graphene (see ref. 20). Here, the electron dynamics is the nonrelativistic Schrödinger-type dynamics. The differential entropy is calculated from our proposed expression (see Eqn. 2 in this article). The slope between the differential entropy and chemical potential gives the µ/D ratio. Through the gate-voltage, we attain the appropriate µ/D value, which plays a vital role in any quantum device. The differential entropy is closely associated with the populated electronic levels, and the chemical potential is responsible for carrier concentration.

Using our entropy-ruled D/µ relation, one can also study the diffusion-mobility relation for various band transport systems (e.g., graphene, silicene, $MoS_2$, etc.). To verify the validity of our method, we extend our calculation from the previous work that was carried by Martin *et al.*, which illustrates the electron-hole puddles in the monolayer graphene system [19]. Here, the density of charges puddles characterized by chemical potential, and the density flux along the electronic levels is directly associated with the diffusion process. In this letter, using Eqn. 3, we extract the differential entropy from different sets of electron density values from this work [19]. In principle, the monolayer graphene is the Dirac-type material (relativistic carrier dynamics), and hence it follows the (2+1)-dimensional charge transport characteristic behaviour. For different gate-voltage-driven chemical potentials, the obtained differential entropy plot is shown Figure 2. It is observed that the slope of this graph provides nonlinear cooperative nature between $h_{eff}$ and η; accordingly, the µ/D ratio is varied at different bias-voltage (chemical potential). Similarly, the differential entropy is extracted from the previous report, which explored the chemical potential-ruled quantum Hall ferromagnetism in bilayer graphene [20]. For our calculation, the relative $h_{eff}$ is numerically obtained using Eqn. 2 from the different sets of gate-voltage-driven electron density values. The chemical potentially enhanced $h_{eff}$ is observed from Figure 3, which also shows the nonlinear synergy between $h_{eff}$ and η. In this extent, the present study suggests that the numerical differentiation of mobility with respect to the diffusion at different chemical potential is a constant for both mono and bilayer graphene systems, which are plotted in Figures 4 and S1.

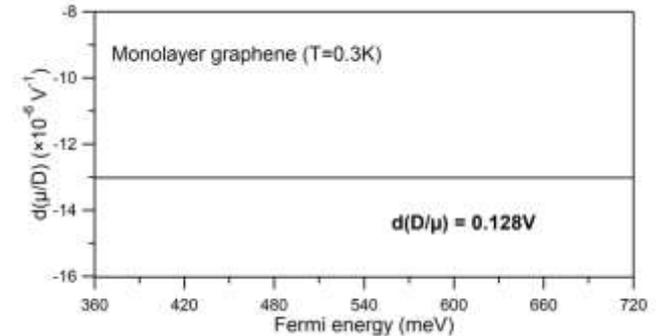

**Figure 4:** The change in D/µ with respect to the chemical potential is a constant value for both the mono and bilayer graphenes. It directly suggests that there is a linear enhancement in D/µ ratio with the chemical potential. The chemical potential naturally incorporates the electronic and thermal parts in the total energy of systems. The electronic energy is mainly correlated with the quantum features, and the classical term is closely connected with temperature. It is observed that for monolayer graphene, the calculated D/µ value is very large due to the relativistic effect. Here, the total energy is equated by kinetic term only. But for other cases (here, bilayer), the carrier dynamics is associated with the kinetic energy as well as other potential terms due to the various typical interactions.

The electronic transport in a monolayer graphene follows the relativistic dynamics, which is characterized by the Dirac equations. In such a Dirac regime, the total energy per particle (average energy) is purely related to kinetic terms only, and other interactions (potential) are negligible [19]. This favours the large D/μ value, which is here dominated by non-interactive motions. The formation of potential energy is expected due to the emergence of interactions (including thermal polaron scattering) while we increase the number of layers. Here, the existence of additional interactions leads to the large possibility of scattering (in the form of potential), which resists the charge transport phenomena via the scattering potential in the electronic media, which is reflected in Figures 4 and S1.

As described in earlier studies, the diffusion-mobility relation gives a direct impact on diode performance, which can be illustrated by current density ($J$)-voltage ($V$) characteristic study via the diode ideality factor ($N_{id}$) [1-3]. For degenerate quantum and nonequilibrium cases, the Navamani-Shockley diode current density equation works well, which includes the quantum correction through the parameters differential entropy and chemical potential [3,9]. In this diode equation, for many-electron system, the total interactions and degeneracy effect (due to applied bias) are accounted by $\eta$ and $h_{eff}$. In this extent, the unified version of Navamani-Shockley equation for different dimensional (1D, 2D and 3D) degenerate quantum devices is

$$J = J_0 \left[ \exp\left( qV \left(\frac{d}{d+2}\right) \frac{\Delta h_S}{\Delta \eta} \right) - 1 \right] \quad (29)$$

As we discussed before, the general form of average energy-Fermi energy relations for 1D, 2D and 3D systems/devices are as follows, $\left.\frac{E}{N}\right|_d = \frac{d}{d+2} E_F$. In principle, the total energy per particle is the chemical potential, which includes all typical interactions along with kinetic energy of the particle. Accordingly, the unified version of Navamani-Shockley diode equation can be expressed as a single form [3],

$$J = J_0 \left[ \exp\left( qV \frac{\Delta h_S}{\Delta E_F} \right) - 1 \right] \equiv J_0 \left[ \exp\left( qV \frac{\Delta h_S}{k_B \Delta T_F} \right) - 1 \right] \quad (30)$$

For disordered molecular systems ($T \geq 150K$), the above Navamani-Shockley diode equation can be reduced as the original Shockley diode equation, $J = J_0 \left[ \exp\left( \frac{qV}{k_B T} \right) - 1 \right]$.

In conclusion, the unified entropy-ruled diffusion-mobility relation is proposed for different dimensional quantum and classical systems. This entropy-ruled μ/D relation is a common for all combinatory systems from molecules to materials, including molecule-material compounds; and it works for entire thermodynamic circumstances in both the zero and non-zero applied bias conditions. We have verified and tested our entropy-ruled relation for thiazolothiazole derivatives **1** and **2**, and for graphene systems. Thiazolothiazole derivative **2** has nearly 3 times of Einstein's relation, i.e., $D/\mu \cong 3k_B T/q$. It is observed that the site energy flux with respect to the effective differential entropy (or vice-versa) along the hopping sites in the molecular systems facilitates the non-equilibrium transport, which is responsible for the deviation in the Einstein relation from its original value, $k_B T/q$. Importantly, the synergy between the effective differential entropy and chemical potential decides the magnitude of D/μ; accordingly, one can explain the validity and limitations of Einstein's relation. For the absence of energy flux $((\Delta E/\Delta h_{eff}) \to 0)$, we naturally retain the Einstein relation from the entropy-ruled method. Besides that the μ/D is verified and tested for the periodic quantum materials via the mono and bilayer graphene systems. In this letter, we have calculated the μ/D values at different chemical potentials using our entropy-ruled relation and the results are discussed. It is noted that the electron dynamics in monolayer is so larger than bialyer graphene, which is hereby verified through D/μ factor. At certain appropriate conditions, the entropy-ruled relation is transformed into the effective mass-based mobility relation, $\mu = q\tau/m^*$. The revisited Boltzmann transport approach with the aid of our entropy-ruled method gives the explicit form of diffusion coefficient for periodic systems. Our unified entropy-ruled analogy for diffusion-mobility is highly solicited, which will hopefully bring a new dimension in semiconductor physics and technology.

**ACKNOWLEDGMENT**

The author is grateful to Prof. Emanuel Tutuc, University of Texas at Austin, United States of America, and Dr. Kayoung Lee for sharing their experimental data of "Science 345, 58 (2014)" to validate our model.

**DEDICATION**

The author dedicated this work to his Godfather Mr. S. MUTHUSAMY (1944-2019), Vadakku thottam, Kavalapatti-624 621, Palani, Tamil Nadu, India for his pivotal mentorship and unconditional support in the author's life.